\documentclass[12pt]{article}
\usepackage[usenames,dvipsnames]{pstricks}
\usepackage{epsfig}
\usepackage{epstopdf}
\usepackage{amsmath}
\usepackage{amssymb}
\usepackage{caption}
\usepackage{subcaption}
\usepackage{amsfonts}

\begin{document}
\title{Probing Non-unitary $CP$ Violation effects in Neutrino Oscillation Experiments }
\author {Surender Verma\thanks{Electronic address: s\_7verma@yahoo.co.in}  and Shankita Bhardwaj\thanks{Electronic address: shankita.bhardwaj982@gmail.com}}

\date{\textit{Department of Physics and Astronomical Science,\\Central University of Himachal Pradesh, Dharamshala 176215, INDIA.}}
\maketitle

\begin{abstract}
In the present work, we have considered minimal unitarity violation(MUV) scheme, to obtain
the general expression for $\nu_{\mu}\rightarrow\nu_{\tau}$ oscillation probability, in vacuum. For this channel, we have investigated the
sensitivities to non-unitary parameters $|\rho_{\mu\tau}|$ and $\omega_{\mu\tau}$ with short baseline(SBL) experiments for normal as well as inverted 
hierarchical neutrino masses. We also check how the sensitivity to non-unitary parameters get modified for $\theta_{23}$ above and below maximality.
We find that the $3\sigma$ sensitivity towards $|\rho_{\mu\tau}|$ is maximum for non-unitary phase $\omega_{\mu\tau}=0$, whereas it is minimum
for $\omega_{\mu\tau}=\pm\pi$ in case of normal hierarchy(NH). However, the sensitivity is minimum at $\omega_{\mu\tau}=0$ and
maximum for $\omega_{\mu\tau}=\pm\pi$ for inverted hierarchy(IH). We observe that for unitary $CP$ phase $\delta=0$ and $\delta=\pi/2$,
the sensitivity to measure non-unitarity remains same in both the cases. We, also, explore wide range of $L/E$
to forecast, in principle, the possibilities to observe $CP$-violation due to unitary($\delta$) and non-unitary($\omega_{\mu\tau}$) phases.
We find that the both phases can be disentangled, in principle, from each other, for the $L/E$ range
less than 200 km/GeV for $\nu_{\mu}\rightarrow\nu_{\tau}$ channel.
\end{abstract} 

\section{Introduction}
The confirmation of neutrino oscillation phenomena, by atmospheric, solar, reactor and accelerator neutrino oscillation
experiments\cite{osc1, osc2, osc3, osc4}, have deepen our knowledge about the neutrino masses and
lepton flavor mixing. Huge progress have been made in extracting the information of
neutrino mixing parameters from the lepton mixing matrix. In three flavor neutrino
mixing paradigm, generally, the determination of neutrino mixing angles and mass squared differences, are based on the assumption that
the lepton mixing matrix is unitary in order to conserve probability. However, in case of generic models including
extra gauge fermionic singlet namely sterile neutrinos or right handed neutrinos, the unitarity of lepton mixing matrix gets
violated and have initiated plethora of research on new physics(NP) beyond Standard Model(SM). Although the total lagrangian containing NP
remain unitary and probability is conserved, but a submatrix of the total mixing matrix may be non-unitary. Moreover, if in any theory the $3\times3$
Pontecorvo-Maki-Nakagava-Sakata(PMNS) mixing matrix considered
to be a part of larger matrix, which comprise of heavy fields(heavier than electroweak
scale), may result in an effective $3\times3$ non-unitary mixing matrix for three light leptonic fields\cite{nu5, nu6, nu7}.
One example, where non-unitarity can be achieved,
is the generic type-I seesaw mechanism\cite{see8, see9, see10, see11}. The non-unitarity effects observed at higher energies(comparable to Grand Unification scale) with type-I seesaw mechanism, the non-unitary effects are small. However, in case of non-minimal seesaw models where
 heavy scale is of the order of TeV, the non-unitarity effects can be appreciable\cite{nm1,nm2,nm3}. In type-I seesaw mechanism, the neutrino masses
are effectively described by unique lepton number violating dimension five Weinberg operator, though the origin of non-unitarity in leptonic mixing matrix is due to the lepton number conserving dimension six operator contributing to the
kinetic terms of the neutrinos\cite{w12, w13}. Furthermore, the significant amount of non-unitarity
can be established with the dimension d = 6 operator, in seesaw mechanism at lower energies.
Beyond SM, NP effects are well addressed by an effective field theory extension named minimal unitarity violation scheme\cite{muv14}. Previously,
under MUV scheme, the non-unitary parameters have been constrained and
connection between MUV and nonstandard interactions have been studied in\cite{muv14, muv15, muv16}.
Moreover, the MUV extensions of SM with significant non-unitarity, can render leptogenesis mechanism, for generating the baryon
asymmetry of the universe, more efficient\cite{muv17}. Recently, constraints on non-unitary parameters have been updated by electroweak precision observable\cite{ew18, ew19}, which also
points towards the non-zero non-unitary parameters. The oscillation channel $\nu_{\mu}\rightarrow\nu_{\tau}$, has been found to have the best sensitivities for observing the non-unitarity,
as discussed in\cite{muv16}. Moreover, the short baseline experiments are excellent probe to search for NP with non-unitary parameter $|\rho_{\mu\tau}|$\cite{tommy}. In the present work, we focussed  on $\nu_{\mu}\rightarrow \nu_{\tau}$ channel to study the non-unitary neutrino mixing. In section 2, we have discussed the
formalism of MUV scheme for parameterising the non-unitary neutrino mixing matrix and to obtain the general expression for oscillation probability
for $\nu_{\mu}\rightarrow\nu_{\tau}$ channel, in vacuum. We have investigated the sensitivities of $\nu_{\mu}\rightarrow\nu_{\tau}$ oscillation  to non-unitary parameters $|\rho_{\mu\tau}|$ and $\omega_{\mu\tau}$ with normal and inverted hierarchical neutrino masses. Also, the effect of $\theta_{23}$, being above or below maximality, on the sensitivities of non-unitary parameters will be discussed. In section 3,
 we explore a very wide range of $L/E$ ratio to forecast in principle, the possibilities to measure the $CP$-violation($CPV$) due to unitary($\delta$) and non-unitary($\omega_{\mu\tau}$) phases. In section 4, we conclude our results.
\section{Neutrino oscillations with non-unitary mixing matrix}
Under MUV extension of standard model, the dimension five operator breaks the electroweak
symmetry by providing masses to the neutrinos. The dimension six operator leads
to the origin of non-unitary mixing matrix by canonical normalization of kinetic terms of neutrinos. 
Non-unitary leptonic mixing matrix can be parameterised as the product of a unitary matrix $U_{0}$ and
a hermitian matrix $H$ \cite{muv14, nu20}, $N=HU_{0}$, where $H\equiv (1+\rho)$ with $\rho=|\rho_{ff'}|e^{-i\omega_{ff'}}$. The flavour eigenstates are
connected to mass eigenstate via non-unitary mixing matrix $N$ as,
\begin{equation}
 \nu_{f}=N_{fi}\nu_{i},
\end{equation}
where $f,i$ are flavor and mass indices, respectively. The oscillation probability for SBL experiments, with unitary and non-unitary contributions can
be expressed as\cite{muv16},
\begin{equation}
P_{ff'}=|\sum_{i=1}^{3}(U_{f'i}e^{-iE_{i}t}U_{fi}^{*})+2 \rho_{ff'}^{*}|^2,
\end{equation}
where, $E_{i}-E_{j}\simeq\Delta m_{ji}^{2}/2E$ and $\Delta m_{ji}^{2}=m_{i}^{2}-m_{j}^{2}$. We have not considered the terms including $\rho$ with subscript other than $ff'$,
as at SBL these terms will not produce significant effects\cite{muv16}, and the maximum sensitivity will be towards the non-unitary parameter $\rho_{ff'}$.  
The exact expression of oscillation probability for $\nu_{\mu}\rightarrow\nu_{\tau}$ channel, in vacuum, is
\begin{eqnarray}
\nonumber
 P_{\mu\tau}=&&4|\rho_{\mu\tau}|^{2}+4c_{12}^{2}c_{23}^4s_{12}^{2}s_{13}^{2}\sin^{2}\left(\Delta_{21}/2\right)+4c_{12}^{2}s_{12}^{2}s_{13}^{2}s_{23}^{4}\sin^{2}\left(\Delta21/2\right)\\
             \nonumber
             &&+4c_{23}^{2}s_{12}^{4}s_{13}^{4}s_{23}^{2}\sin^{2}\left(\Delta_{21}/2\right)-8c_{12}^{2}c_{23}^{2}s_{12}^{2}s_{13}^{2}s_{23}^{2}\sin^{2}\left(\Delta_{21}/2\right)\\
             \nonumber
             &&-8c_{12}^{2}c_{23}|\rho_{\mu\tau}|s_{23}\sin\left(\Delta_{21}/2\right)\cos\left(\omega_{\mu\tau}-\Delta_{21}/2\right)+4c_{12}^{4}c_{23}^{2}s_{23}^{2}\sin^{2}\left(\Delta_{21}/2\right)\\
           \nonumber
               &&-8c_{12}^{2}c_{23}^{2}s_{12}^{2}s_{13}^{2}s_{23}^{2}\sin^{2}\left(\Delta_{21}/2\right)\cos2\delta+4c_{13}^{4}c_{23}^{2}s_{23}^{2}\sin^{2}\left(\Delta_{31}/2\right)\\
            \nonumber
 &&+8c_{23}|\rho_{\mu\tau}|s_{12}^{2}s_{13}^{2}s_{23}\sin\left(\Delta_{21}/2\right)\cos\left(\omega_{\mu\tau}-\Delta_{21}/2\right)\\
   \nonumber
            &&-8c_{12}c_{23}^{2}|\rho_{\mu\tau}|s_{12}s_{13}\cos\delta\sin\left(\Delta_{21}/2\right)\cos\left(\omega_{\mu\tau}-\Delta_{21}/2\right)\\
            \nonumber
            &&+8c_{12}|\rho_{\mu\tau}|s_{12}s_{13}s_{23}^{2}\cos\left(\Delta_{21}/2\right)\sin\left(\Delta_{21}/2\right)\cos\left(\delta+\omega_{\mu\tau}\right)\\
            \nonumber
            &&+8c_{12}|\rho_{\mu\tau}|s_{12}s_{13}s_{23}^{2}\sin^{2}\left(\Delta_{21}/2\right)\sin\left(\delta+\omega_{\mu\tau}\right)\\
            \nonumber
            &&+8c_{12}^{3}c_{23}^{3}s_{12}s_{13}s_{23}\cos\delta\sin^{2}\left(\Delta_{21}/2\right)-8c_{12}^{3}c_{23}s_{12}s_{13}s_{23}^{3}\cos\delta\sin^{2}\left(\Delta_{21}/2\right)\\
            \nonumber
            &&-8c_{12}c_{23}^{3}s_{12}^{3}s_{13}^{3}s_{23}\cos\delta\sin^{2}\left(\Delta_{21}/2\right)+8c_{12}c_{23}s_{12}^{3}s_{13}^{3}s_{23}^{3}\cos\delta\sin^{2}\left(\Delta_{21}/2\right)\\
            \nonumber
            &&+8c_{13}^{2}c_{23}|\rho_{\mu\tau}|s_{23}\sin\left(\Delta_{31}/2\right)\cos\left(\omega_{\mu\tau}-\Delta_{31}/2\right)-8c_{12}^{2}c_{13}^{2}c_{23}^{2}s_{23}^{2}A\\
            \nonumber
            &&+8c_{13}^{2}c_{23}^{2}s_{12}^{2}s_{13}^{2}s_{23}^{2}A-8c_{12}c_{13}^{2}c_{23}^{3}s_{12}s_{13}s_{23}\cos\delta A+8c_{12}c_{13}^{2}c_{23}s_{12}s_{13}s_{23}^{3}\cos\delta A\\
           \nonumber
            &&-8c_{12}c_{13}^{2}c_{23}^{3}s_{12}s_{13}s_{23}\sin\delta B-8c_{12}c_{13}^{2}c_{23}s_{12}s_{13}s_{23}^{3}\sin\delta B\\
          &&-8c_{12}c_{23}^{2}|\rho_{\mu\tau}|s_{12}s_{13}\sin\delta\sin\left(\Delta_{21}/2\right)\sin\left(\omega_{\mu\tau}-\Delta_{21}/2\right),
 \end{eqnarray}

where $c_{ij}=\cos\theta_{ij}$, $s_{ij}=\sin\theta_{ij}$, $\Delta_{ji}=1.27\frac{\Delta m_{ji}^2L}{E}$, $L$ is the
baseline length(km), $E$ is the neutrino beam energy(GeV), $A=\sin\left(\Delta_{21}/2\right)\sin\left(\Delta_{31}/2\right)\cos\left(\Delta_{32}/2\right)$ and $B=\sin\left(\Delta_{21}/2\right)\sin\left(\Delta_{31}/2\right)\sin\left(\Delta_{32}/2\right)$. Using trigonometric identities, Eqn. (3) can be written as
\begin{eqnarray}
\nonumber
 P_{\mu\tau}=&&4|\rho_{\mu\tau}|^{2}+\bigl[\sin^{2}2\theta_{23}c_{13}^{4}\bigr]\sin^{2}\left(\Delta_{31}/2\right)+\bigl[\sin^{2}2\theta_{12}\left(c_{23}^{4}+s_{23}^{4}\right)\\
            \nonumber
            &&+\sin^{2}2\theta_{23}\left(s_{12}^{4}s_{13}^{4}+c_{12}^{4}\right)-s_{13}^{2}\sin^{2}2\theta_{23}\sin^{2}2\theta_{12}\cos^{2}\delta\\
            \nonumber
            &&+s_{13}\sin2\theta_{12}\sin4\theta_{23}\cos\delta\left(c_{12}^{2}-s_{12}^{2}s_{13}^{2}\right)\bigr]\sin^{2}\left(\Delta_{21}/2\right)\\    
      \nonumber
            &&+\bigl[-2c_{13}^{2}\sin^{2}2\theta_{23}\left(c_{12}^{2}-s_{12}^{2}s_{13}^{2}\right)-(1/2)c_{13}\sin2\theta_{12}\sin2\theta_{13}\sin4\theta_{23}\\
            \nonumber
            &&\cos\delta\cos(\Delta_{32}/2)-c_{13}\sin2\theta_{12}\sin2\theta_{13}\sin2\theta_{23}\sin\delta\sin(\Delta_{32}/2)\bigr]\sin(\Delta_{21}/2)\\
            \nonumber
            &&\sin(\Delta_{31}/2)+\bigl[4|\rho_{\mu\tau}|s_{12}^{2}s_{13}^{2}\sin2\theta_{23}\cos\left(\omega_{\mu\tau}-\Delta_{21}/2\right)-4c_{23}^2s_{13}|\rho_{\mu\tau}|\sin2\theta_{12}\\
            \nonumber
            &&\cos\left(\delta-\omega_{\mu\tau}+\Delta_{21}/2\right)+4s_{23}^{2}s_{13}|\rho_{\mu\tau}|\sin2\theta_{12}\cos\left(\delta+\omega_{\mu\tau}-\Delta_{21}/2\right)\bigr]\\
           &&\sin\left(\Delta_{21}/2\right)+\bigl[4c_{13}^2\sin2\theta_{23}|\rho_{\mu\tau}|\cos\left(\omega_{\mu\tau}-\Delta_{31}/2\right)\bigr]\sin\left(\Delta_{31}/2\right).
\end{eqnarray}
      
 The first term in the Eqn.(4) represents zero distance effects(no $L$ dependence). The coefficients of $\sin^{2}\left(\Delta_{31}/2\right)$, $\sin^{2}(\Delta_{21}/2)$ and $\sin(\Delta_{21}/2)\sin(\Delta_{31}/2)$ in second, third and fourth terms, respectively, depends on unitary $CP$ phase $\delta$ only. Whereas, the coefficients of $\sin(\Delta_{21}/2)$ and $\sin(\Delta_{31}/2)$ in fifth and sixth terms, respectively, have dependence on both unitary($\delta$) as well as non-unitary($\omega_{\mu\tau}$) $CP$ phases.
\begin{figure}[htp]
\centering
\parbox{6cm}{
\includegraphics[width=5cm]{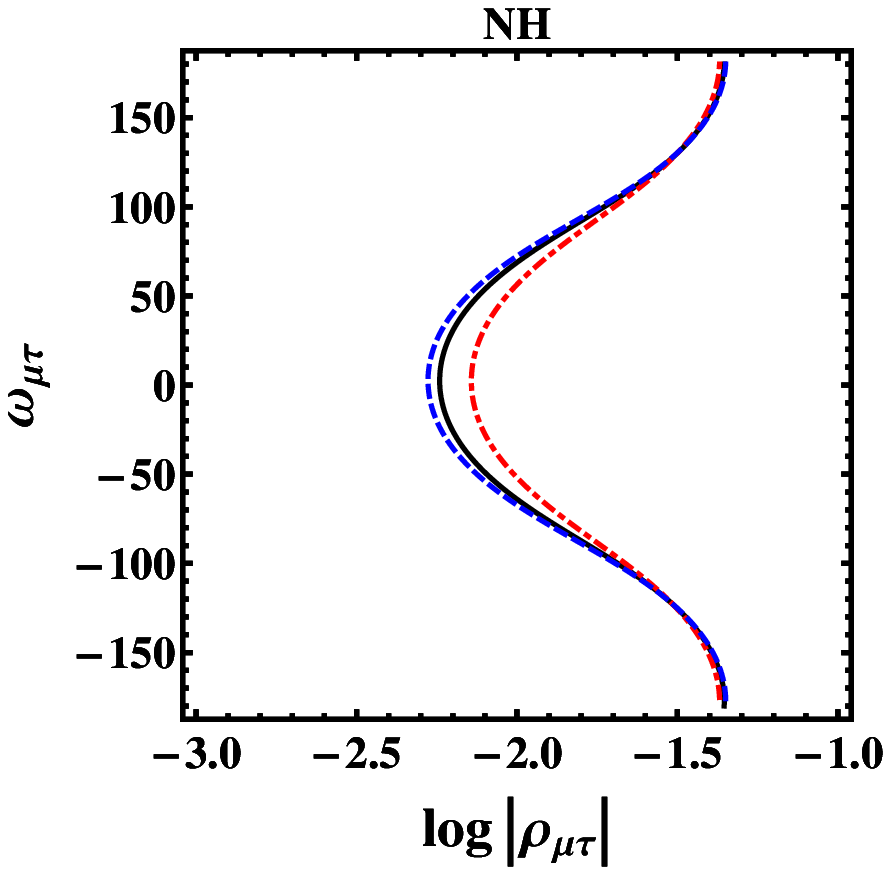}}
\qquad
\begin{minipage}{5cm}
\includegraphics[width=5cm]{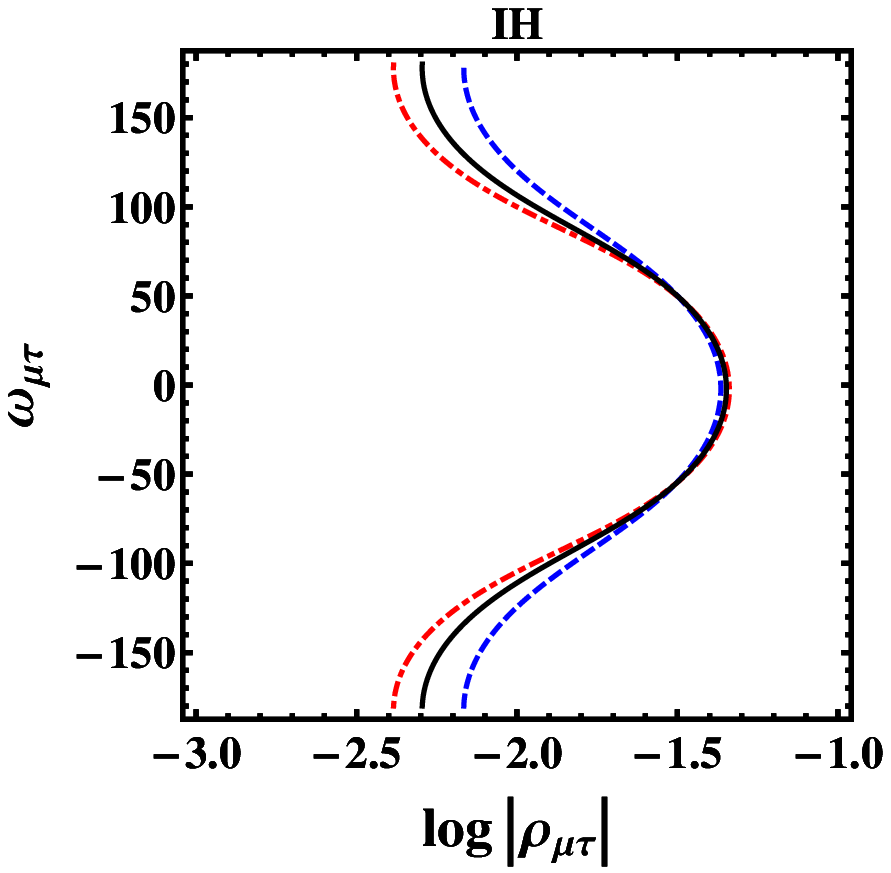}

\end{minipage}
\centering
\parbox{6cm}{
\includegraphics[width=5cm]{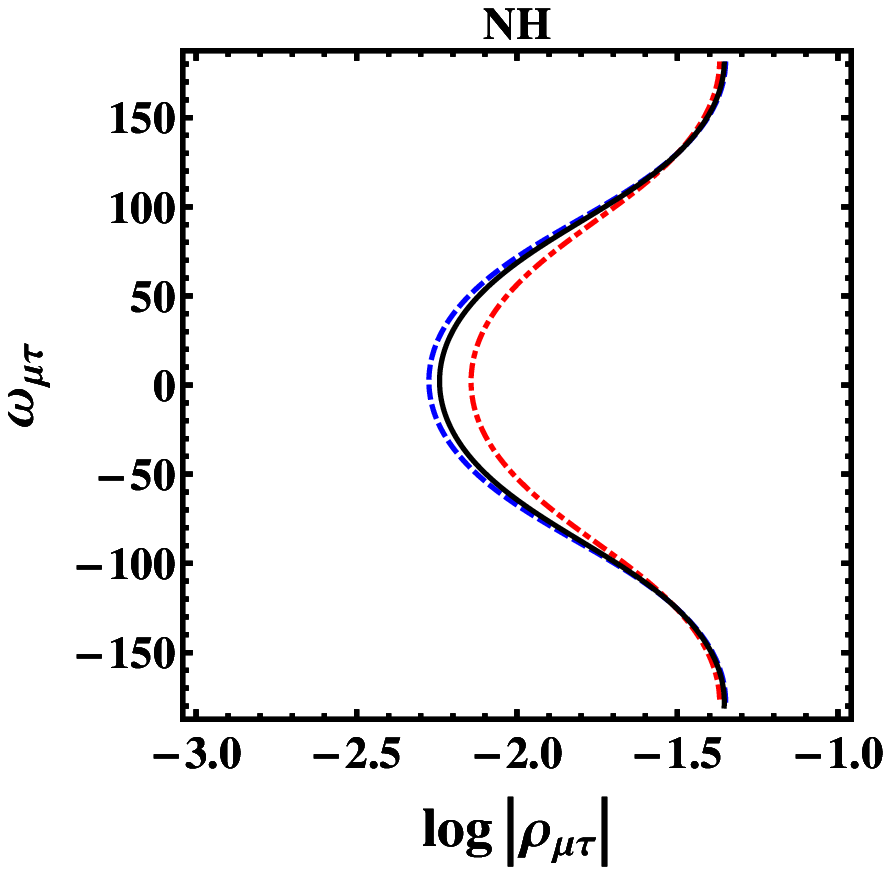}}
\qquad
\begin{minipage}{5cm}
\includegraphics[width=5cm]{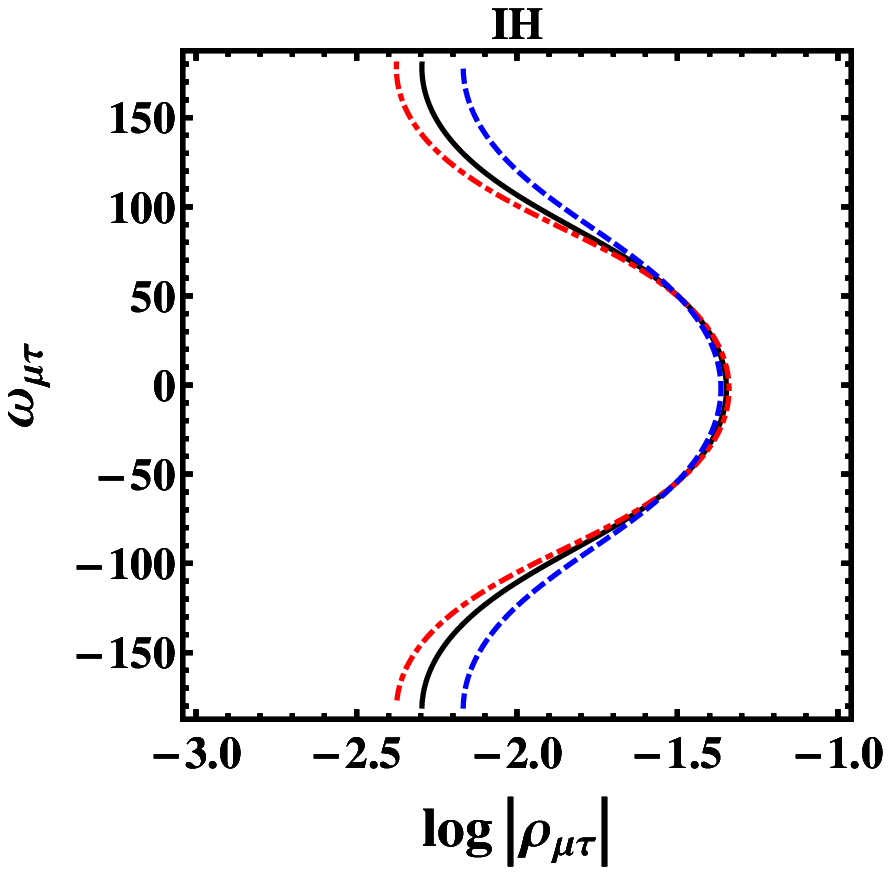}
\end{minipage}
\caption{$3\sigma$ contour plots for the Normal Hierarchical(left) and Inverted Hierarchical(right) neutrino mass spectrum. First(second) row
represents the contour plots for $\delta=0$($\delta=\pi/2$). The dashed(dotdashed) and solid lines represents $3\sigma$ upper(lower) and bestfit values.}
\end{figure}

Using global data for neutrino mixing parameters\cite{data} to study short baseline experiments with baseline $L=130$ km between neutrino factory source and OPERA-like near detector(which is similar to the distance between CERN and FREJUS). We assume neutrino beam energy $E_{\nu}=5$ GeV and obtain $3\sigma$ sensitivity contours for non-unitary parameters. This baseline have been studied previously in various studies\cite{muv16, base22} with different energy beams. It is evident from Fig.(1) that
sensitivity to $|\rho_{\mu\tau}|$ is maximum around $\omega_{\mu\tau}=0$ and minimum at $\omega_{\mu\tau}=\pm\pi$ for normal hierarchical neutrino masses. This behaviour can be comprehended from the last two terms of Eqn. (4) containing $\omega_{\mu\tau}$ and $|\rho_{\mu\tau}|$. For inverted hierarchical neutrino masses the situation get reversed and sensitivity to $|\rho_{\mu\tau}|$ is maximum around $\omega_{\mu\tau}=\pm\pi$ and minimum at $\omega_{\mu\tau}=0$. In case of IH, when $\omega_{\mu\tau}=0$, the two terms will give positive contribution
and two will give negative contribution to the probability. The contribution from $\Delta_{31}$ will be significantly dominant over $\Delta_{21}$ and
hence the dominant contributions are from negative terms and net sensitivity for non-unitary parameter $\rho_{\mu\tau}$ will be minimum in this case. Further, for $\omega_{\mu\tau}=\pm\pi$, there will be two terms with negative and two terms with positive contribution as similar to previous case, but here the term containing $\Delta_{31}$ with positive contribution is dominant over the contribution from $\Delta_{21}$ resulting in maximum sensitivity at $\omega_{\mu\tau}=\pm\pi$. Moreover, the contours have reflection symmetry about $\omega_{\mu\tau}=0$ for both normal as well as inverted hierarchical neutrino masses because of cosine dependence of $\omega_{\mu\tau}$. The possibility of $\theta_{23}$ being above or below maximality have distinguishing implications for sensitivities in normal and inverted hierarchies of neutrino masses. Sensitivity to $|\rho_{\mu\tau}|$ is maximum for $\theta_{23}$ above maximal for NH whereas it is maximum for $\theta_{23}$ below maximality for IH. Also, for $\omega_{\mu\tau}=\pm\pi$(NH) and $\omega_{\mu\tau}=0$(IH), we cannot have information about the octant of $\theta_{23}$. So observation of non-unitarity effects in neutrino oscillations have profound implications on determining the neutrino mass hierarchy and octant of $\theta_{23}$.

\section{Disentangling unitary and non-unitary $CPV$ effects}
In order to investigate intermixing, and prospects for possible detection, of the unitary and non-unitary $CP$ phases in $P_{\mu\tau}$,
the oscillation probability in Eqn.(3) can be seceded in four components
\begin{equation}
 P_{\mu\tau}=P_{0}+P_{\delta}+P_{\omega_{\mu\tau}}+P_{\delta,\omega_{\mu\tau}}.
\end{equation}

First component, $P_0$, is independent of the $CP$ phases $\delta$ and $\omega_{\mu\tau}$ and will, further,  be ignored in the analysis
\begin{eqnarray}
\nonumber
 P_{0}=&&4|\rho_{\mu\tau}|^{2}+s_{13}^{2}\left(c_{23}^4+s_{23}^4\right)\sin^2 2\theta_{12}\sin^{2}\left(\Delta21/2\right)\\
            \nonumber
            &&-(1/2)s_{13}^{2}\sin^2 2\theta_{23}\sin^2 2\theta_{12}\sin^{2}\left(\Delta_{21}/2\right)\\
            \nonumber
            &&+\left(s_{12}^{4}s_{13}^{4}+c_{12}^4\right)\sin^2 2\theta_{23}\sin^{2}\left(\Delta_{21}/2\right)\\
            \nonumber
            &&-2c_{13}^2\left(c_{12}^{2}-s_{12}^{2}s_{13}^{2}\right)A\sin^2 2\theta_{23}\\
            &&+c_{13}^{4}\sin^{2}2\theta_{23}\sin^{2}\left(\Delta_{31}/2\right),\\
            \nonumber
            \end{eqnarray}

Second (third) component, $P_{\delta}$ ($P_{\omega_{\mu\tau}}$) depends on unitary phase (non-unitary phase), $\delta$ ($\omega_{\mu\tau}$). The last component, $P_{\delta,\omega_{\mu\tau}}$ has mixed dependence on the $CP$ phases. These different components of the oscillation probability $P_{\mu\tau}$($P_\delta, P_{\omega_{\mu\tau}}$ and $P_{\delta,\omega_{\mu\tau}}$) can be written as

\begin{eqnarray}
\nonumber
P_{\delta}=&&-(1/2)s_{13}^{2}\sin^{2}2\theta_{12}\sin^{2}2\theta_{23}\cos 2\delta\sin^2\left(\Delta_{21}/2\right)\\
 \nonumber
            &&+s_{13}\left(c_{12}^{2}-s_{12}^{2}s_{13}^{2}\right)\sin 4\theta_{23}\sin 2\theta_{12}\cos\delta\sin^{2}\left(\Delta_{21}/2\right)\\
            &&-c_{13}\sin 2\theta_{23}\sin 2\theta_{12}\sin 2\theta_{13}\left(A\cos 2\theta_{23}\cos\delta+B\sin\delta\right),  \\
         \nonumber
            \\
 \nonumber
P_{\omega_{\mu\tau}}=&&-4|\rho_{\mu\tau}|\left(c_{12}^2-s_{12}^2s_{13}^2\right)\sin 2\theta_{23}\sin\left(\Delta_{21}/2\right)\cos\left(\omega_{\mu\tau}-\Delta_{21}/2\right)\\
                    &&+4c_{13}^{2}|\rho_{\mu\tau}|\sin 2\theta_{23}\sin\left(\Delta_{31}/2\right)\cos\left(\omega_{\mu\tau}-\Delta_{31}/2\right),\\
 \nonumber
 \\
 \nonumber 
 \\
 \nonumber
P_{\delta,\omega_{\mu\tau}}=&&-4s_{13}c_{23}^{2}|\rho_{\mu\tau}|\sin 2\theta_{12}\sin\left(\Delta_{21}/2\right)\cos\left(\Delta_{21}/2+\delta-\omega_{\mu\tau}\right)\\
                              &&+4s_{13}s_{23}^{2}|\rho_{\mu\tau}|\sin 2\theta_{12}\sin\left(\Delta_{21}/2\right)\cos\left(\Delta_{21}/2-\delta-\omega_{\mu\tau}\right).
                          \end{eqnarray}

\begin{figure}[htp]
\centering
\includegraphics[width=0.5\textwidth]{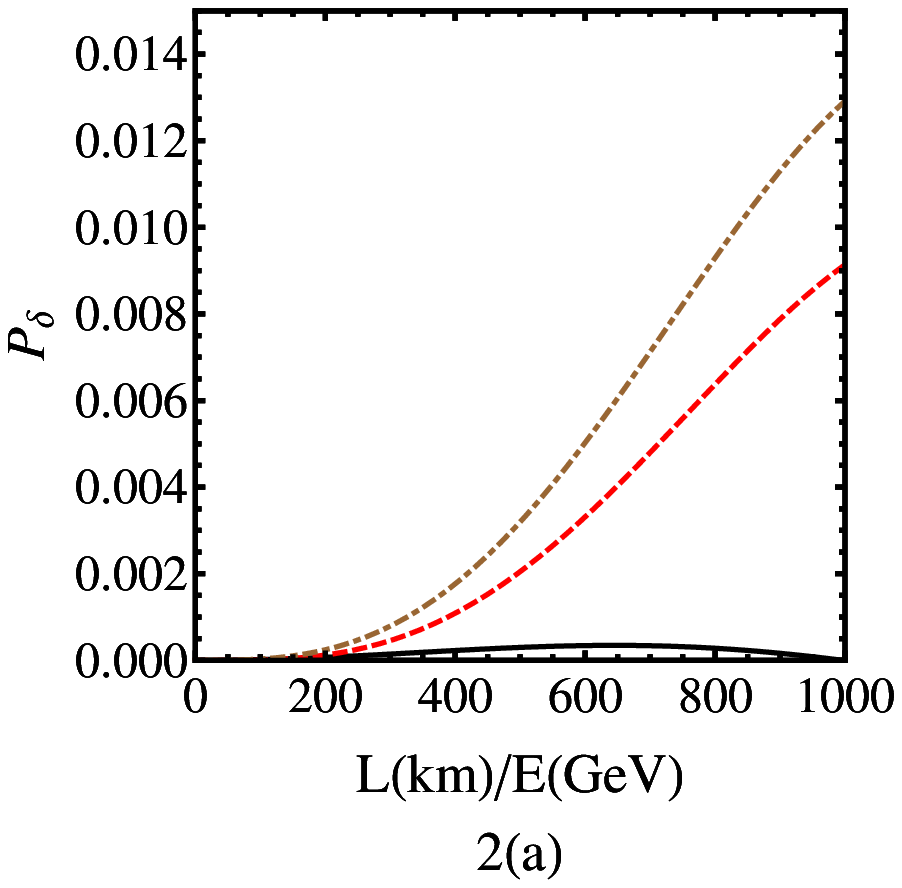}\hfill
\includegraphics[width=0.5\textwidth]{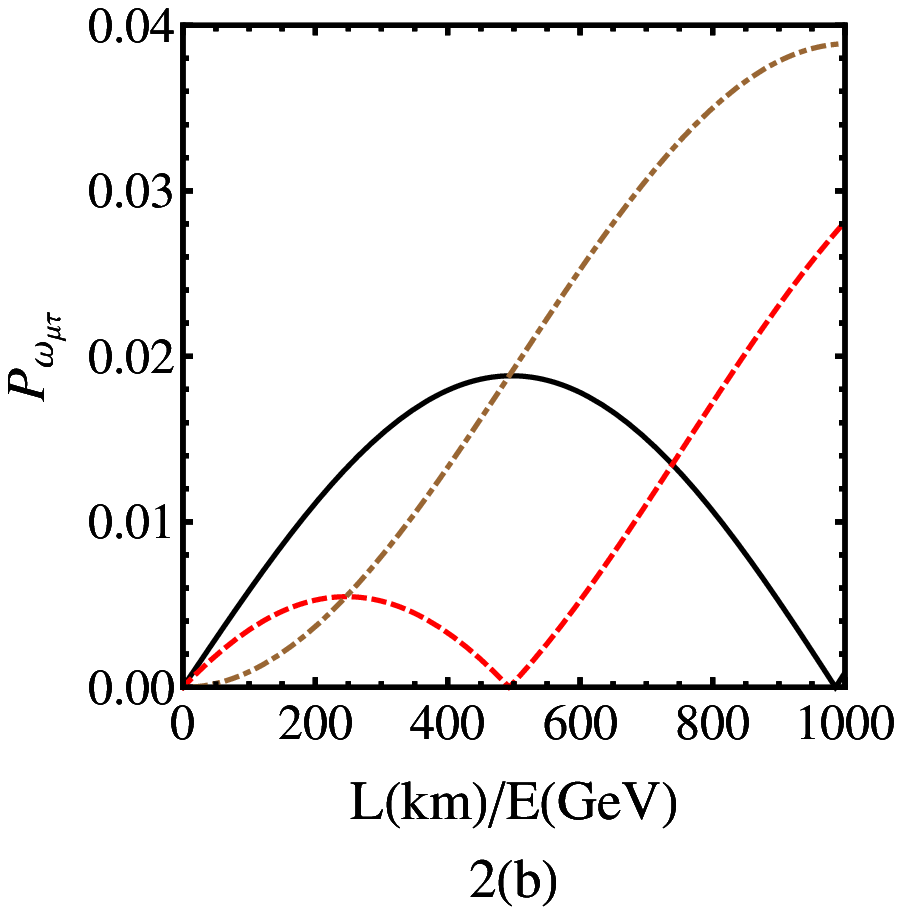}\hfill
\includegraphics[width=0.5\textwidth]{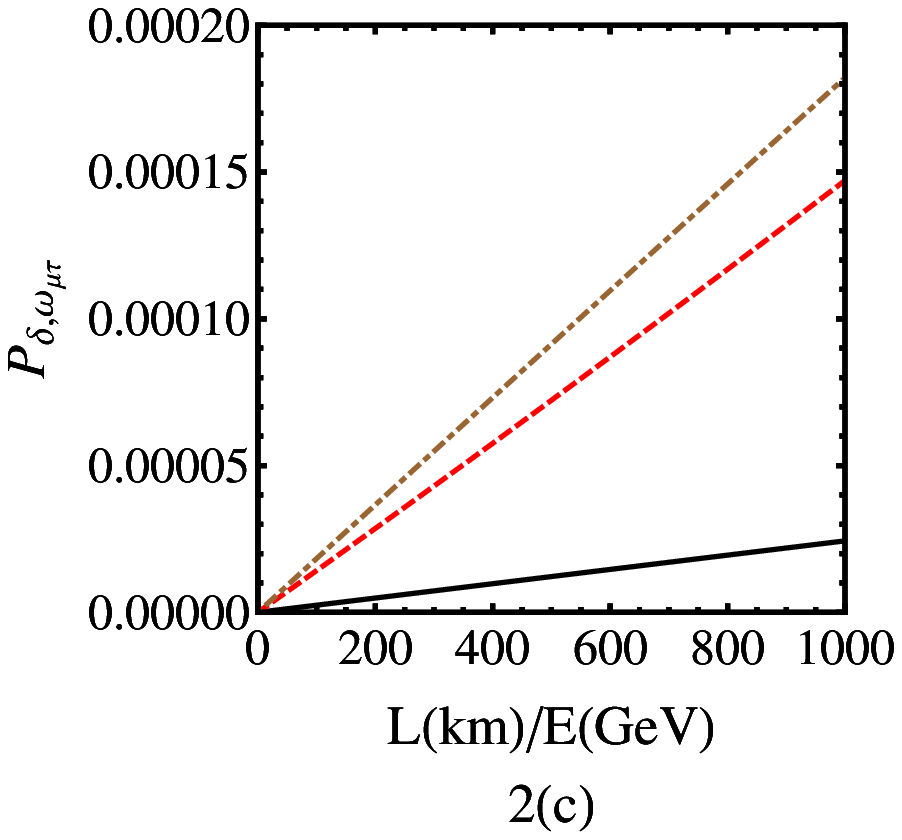}

\caption{$P_{\delta},P_{\omega_{\mu\tau}}$ and $P_{\delta,\omega_{\mu\tau}}$ as a function of $L/E$(km/GeV). 
In Fig.2(a)(Fig.2(b)), the solid line has been obtained for $\delta=0$($\omega_{\mu\tau}=0$), dashed for $\delta=\pi/4$($\omega_{\mu\tau}=\pi/4$) and dotdashed for $\delta=\pi/2$($\omega_{\mu\tau}=\pi/2$). In Fig.2(c), solid line has been plotted for $\delta=0$ and $ \omega_{\mu\tau}=0$, dashed line for $\delta=\pi/4$ and $\omega_{\mu\tau}=\pi/4$ and dotdashed line for $\delta=\pi/2$ and $\omega_{\mu\tau}=\pi/2$.}
\end{figure}

In Fig.(2), we have plotted $P_{\delta}, P_{\omega_{\mu\tau}}$ and $P_{\delta,\omega_{\mu\tau}}$ with $L/E$. We can see from Fig.2(a) that $P_{\delta}$ is negligibly small for $L/E\le 200$ km/GeV. For this range of $L/E$, $P_{\omega_{\mu\tau}}$ is dominant component in the total oscillation probability $P_{\mu\tau}$ irrespective of the value of non-unitary parameter, $\omega_{\mu\tau}$(Fig. 2(b)). Fig. 2(b) has been obtained for three representative values of $\omega_{\mu\tau}=0,\pi/4$ and $\pi/2$. Also, the contribution of $P_{\delta,\omega_{\mu\tau}}$ is $\mathcal{O}(10^{-5})$. Thus, the oscillation experiments with $L/E\le 200$ km/GeV have, in principle, bright prospects for the investigation of the $CP$ effects, in $P_{\mu\tau}$, due to non-unitary $CP$ phase. However, it will require combination of several oscillation experiments with different baselines and energy spectrum.

\section{Conclusions}

In conclusion, we have investigated the sensitivities of $\nu_{\mu}\rightarrow\nu_{\tau}$ oscillation  to non-unitary parameters $|\rho_{\mu\tau}|$ and $\omega_{\mu\tau}$ with normal and inverted hierarchical neutrino masses. Also, the effect of $\theta_{23}$, being above or below maximality, on the sensitivities of non-unitary parameters are discussed. We find that the sensitivity to $|\rho_{\mu\tau}|$ is maximum around $\omega_{\mu\tau}=0$ and minimum at $\omega_{\mu\tau}=\pm\pi$ for normal hierarchical neutrino masses. This behaviour can be comprehended from the last two terms of Eqn. (4) containing $\omega_{\mu\tau}$ and $|\rho_{\mu\tau}|$. For inverted hierarchical neutrino masses the situation get reversed and sensitivity to $|\rho_{\mu\tau}|$ is maximum around $\omega_{\mu\tau}=\pm\pi$ and minimum at $\omega_{\mu\tau}=0$. The probability contours have reflection symmetry about $\omega_{\mu\tau}=0$ for both normal as well as inverted hierarchical neutrino masses because of cosine dependence of $\omega_{\mu\tau}$. The possibility of $\theta_{23}$ being above or below maximality have distinguishing implications for sensitivities in normal and inverted hierarchies of neutrino masses. Sensitivity to $|\rho_{\mu\tau}|$ is maximum for $\theta_{23}$ above maximal for NH whereas it is maximum for $\theta_{23}$ below maximality for IH. Also, for $\omega_{\mu\tau}=\pm\pi$(NH) and $\omega_{\mu\tau}=0$(IH), we cannot have information about the octant of $\theta_{23}$. So observation of non-unitarity effects in neutrino oscillations have profound implications on determining the neutrino mass hierarchy and octant of $\theta_{23}$. Furthermore, we illustrated that the sensitivity remain same for unitary $CP$ phase $\delta=0$ and $\delta=\pi/2$,
as the contribution from unitary phase is so small to be observed at this baseline. On exploring a very broad spectrum of $L/E$ (km/GeV) ratio,
we find that non-unitarity $CP$ effects in the neutrino mixing may be explored in oscillation experiments having $L/E\leq200$ km/GeV. However, it will require data from combination of oscillation experiments with different baselines or making use of wide energy spectrum.

\vspace{1cm}
\textbf{\Large{Acknowledgements}}\\
S. V. acknowledges the  financial support provided by University Grants Commission (UGC)-Basic Science
Research(BSR), Government of India vide Grant No. F.20-2(03)/2013(BSR). S. B. acknowledges the financial 
support provided by the Central University of Himachal Pradesh.

\end{document}